\documentclass[a4paper,11pt]{article}
\usepackage{jinstpub} % for details on the use of the package, please
\usepackage{siunitx}
\usepackage{amsmath}
\usepackage{subfigure}
\usepackage{multirow}
\usepackage{hyperref}

\hypersetup{colorlinks,breaklinks,
  linkcolor=black,citecolor=blue,
  filecolor=blue,urlcolor=blue,
  pdfpagemode=UseNone
}
\usepackage{lineno}% \linenumbers to turn on

\usepackage{etoolbox} %% <- for \cspreto, \csappto
\newcommand*\linenomathpatch[1]{%
  \cspreto{#1}{\linenomath}%
  \cspreto{#1*}{\linenomath}%
  \csappto{end#1}{\endlinenomath}%
  \csappto{end#1*}{\endlinenomath}%
}
\linenomathpatch{equation}

\hypersetup{breaklinks}
\graphicspath{{./}{figure/}} %wm% searching path

\usepackage{siunitx}
 \usepackage{hyperref}
 \hypersetup{colorlinks,breaklinks,
   linkcolor=blue,citecolor=blue,
   filecolor=blue,urlcolor=blue,
 }

\usepackage{caption}
\newcommand*{\refcite}[1]{ref.~\cite{#1}}
\newcommand*{\tab}[1]{Table~\ref{tab:#1}}
\newcommand*{\fig}[1]{Figure~\ref{fig:#1}}
\newcommand*{\sect}[1]{Section~\ref{sect:#1}}

\newcommand{\X}[3][]{\ensuremath{{}_{#1}^{#3}\text{{#2}}}\xspace}% \X[Z]{name}{A} element

\newcommand{\supixi}{\textsc{Supix}-1\xspace}

\title{\boldmath Radiation hardness study on a CMOS pixel sensor for charged particle tracking}

\author[a]{L. Li,}%%\note{Corresponding author.}}
\author[a, 1]{L. Zhang,\note{Corresponding author.}}
\author[a]{J. N. Dong,}
\author[b]{J. Liu,}
\author[a, 1]{M. Wang}%%\note{Corresponding author.}}

\affiliation{Institute of Frontier and Interdisciplinary Science and
  Key Laboratory of Particle Physics and Particle Irradiation,
  Shandong University,\\Qingdao, 266237, Shandong, China}
 \affiliation[b]{Department of Physics, University of Liverpool,
  Liverpool, L693BX, United Kingdom}
\emailAdd{mwang@sdu.edu.cn}
\emailAdd{zhang.l@sdu.edu.cn}

\abstract{
A CMOS pixel sensor, named \supixi, is developed for a pixelated silicon tracker for the Circular 
Electron-Positron Collider (CEPC) project. The sensor, consisted of nine sectors varying in pixel sizes, 
diode sizes and geometries, is fabricated with a 180\, nm CMOS Image Sensor (CIS) process to study the 
particle detection performance of enlarged pixels. 
In this work, the radiation-induced effects on the charge collection of the sensor under the fluence of 
$1\times10^{13}~1\,\text{MeV}~\text{n}_\text{eq}/\text{cm}^2$ are studied by the measurements with 
the radioactive source of \X{Fe}{55} and the Technology Computer Aided Design (TCAD) simulations, 
since the radiation hardness of $6.8\times10^{12}~1\,\text{MeV}~\text{n}_\text{eq}/\text{cm}^2$ per year 
for Non-Ionizing Energy Loss (NIEL) effects is required. 
In measurements,  the sensor gain has been calibrated using the k-$\alpha$ peak of \X{Fe}{55} before and after 
irradiation. The pixel-wise equivalent noise charge (ENC), charge collection efficiency (CCE) and signal-to-noise 
ratio (SNR) were evaluated.
The radiation-induced effects on cluster properties are studied through a self-developed reconstruction algorithm. 
In TCAD simulations, charge collections in $5\times5$ pixel matrixes for two typical impinging cases of incident 
particles were simulated with and without irradiation.
Both measurements and simulations indicate that enlarged pixels with area of \SI{21x84}{\um}, though suffering greater 
loss on sensor performance than small pixels do, still have satisfactory noise and charge collection performance after 
irradiation for particle tracking in the upcoming collider detectors.
   }

\keywords{CMOS imagers, Particle tracking detector and programs, TCAD, Radiation-hard detectors}

\arxivnumber{1234.56789} % only if you have one

\begin{document}
\maketitle
\flushbottom

%___________________________________________________________________________________________________________________
\section{Introduction}
\label{sect:a}

\begin{figure}
  \centering
  \subfigure[]{
    \includegraphics[width=0.9\textwidth,origin=c]{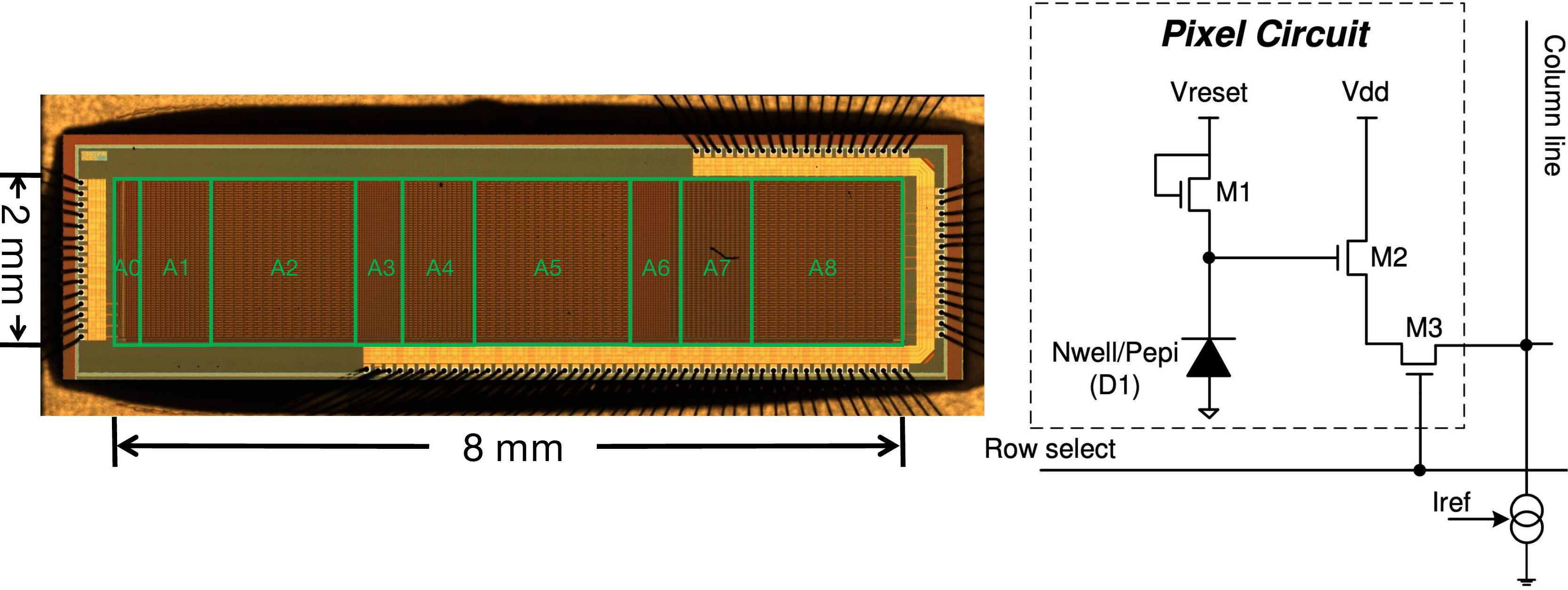}
    \label{fig:chip:a}
  }
  
  \subfigure[]{
    \includegraphics[width=0.35\textwidth,origin=c]{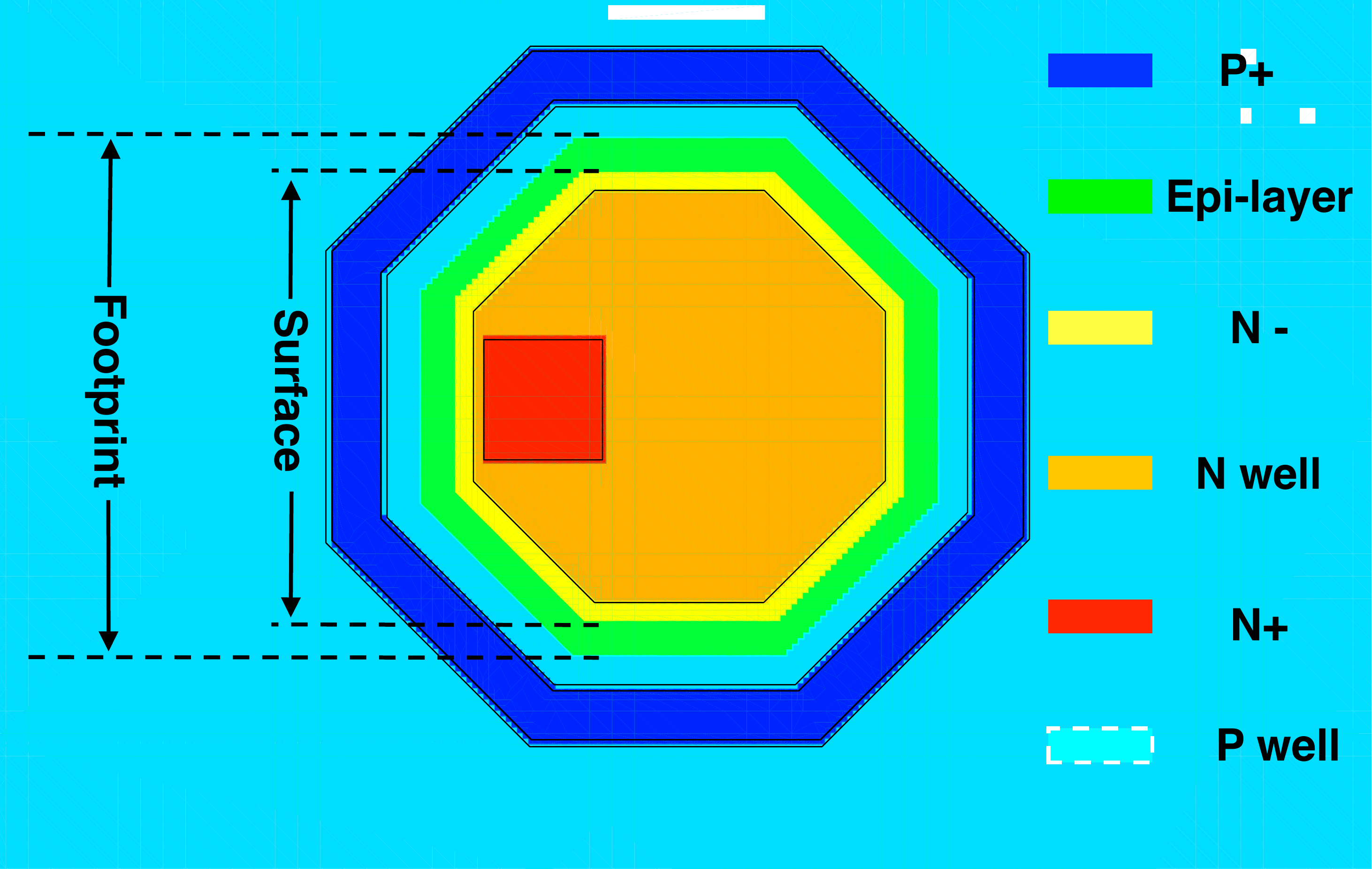}
    \label{fig:chip:b}
  }
  \subfigure[]{
  \includegraphics[width=0.6\textwidth, origin=c]{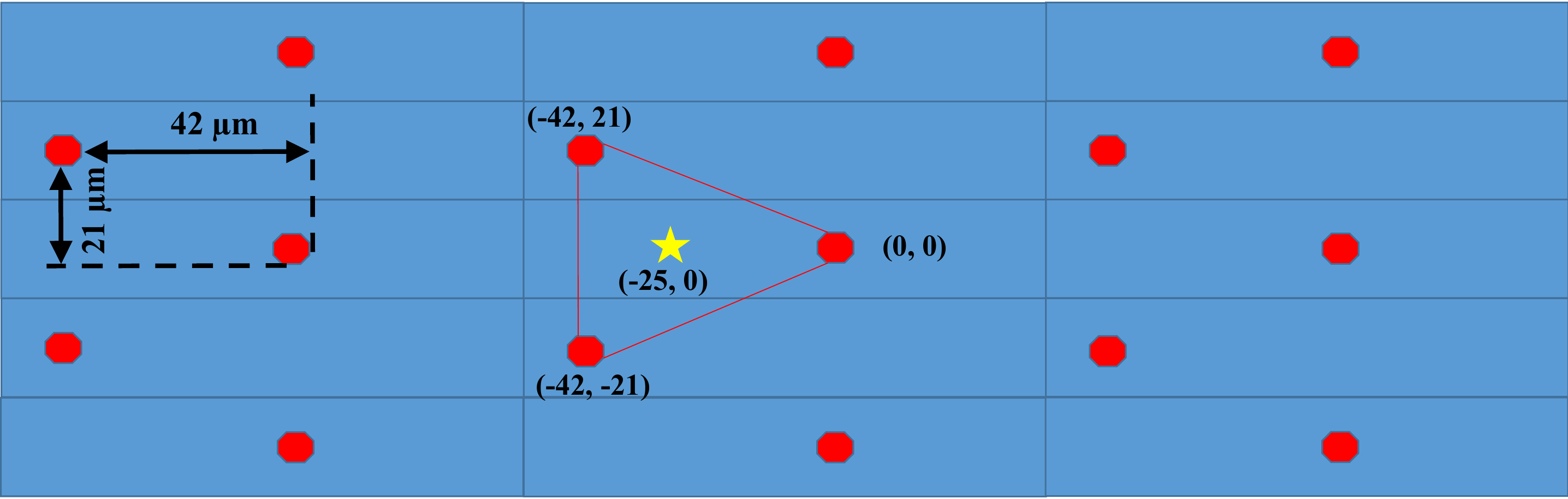}
  \label{fig:chip:c}
  }
  
  \caption{Design of the \supixi sensor: (a) a photograph and the schematic 
  	of in-pixel circuit, (b) the geometry of a sensor diode and (c) the staggered 
	arrangement of diodes in sectors with the largest x-pitch.}
\end{figure}

A CMOS pixel sensor, named Shandong University PIXel~(\supixi), is developed for silicon tracker of CEPC\cite{cepcacc, cepcphys} 
for precision tracking.  
Shown in \fig{chip:a}, the sensor chip covers a sensitive area of $2\times8~\text{mm}^{2}$ containing 
9 sectors (A0-A8). Each sector contains 64 rows and 16 columns of pixels.
%
%According to the general condition of 
%the single point resolution, $\sigma_\text{sp} <  7\,\mu\text{m}$, for CEPC silicon tracker on the transverse plane with 
%regard to the magnetic field\cite{cepcphys}, relatively larger pixels of \SI{21x84}{\um}, comparing to the ULTIMATE sensor 
%(\SI{20.7x20.7}{\um})\cite{starPXL} and the
%ALPIDE sensor (\SI{27x29}{\um})\cite{alpide} which aim for the vertex detector, are realized in \supixi regardless of 
%the charge collection. 
%
%In order to investigate the interplay of the pixel pitch and the diode geometry on charge collection and justify the 
%feasibility of relatively large pixels, 
Three pixel pitches and six sets of diode geometries are implemented in the
sensor, whereas five sectors are functional in the test (A0, A2, A5, A7 and A8). The sensitive area, the pixel pitch 
and the diode geometry for corresponding sectors are listed in \tab{result}. The diode surface is defined as the area 
of N-well and the diode footprint is the area surrounded by P-well, shown in \fig{chip:b}. Additionally, a staggered 
pixel design, shown in \fig{chip:c}, is implemented in sectors with large x-pitch ($84~\mu$m) to enhance the charge 
collection.  
%
%A 3T architecture for pixel circuit is employed for readout. Signal charge is collected through N-well/P-epitaxial layer 
%diode~(D1). Analog signal for each sector is processed frame by frame in rolling shutter mode individually, with a 
%2~MHz basic clock. 
Details of the pixel design can be found in \refcite{supix1tcad}. 

The radiation hardness of bulk damage on sensors, scaled with the NIEL\cite{pixel_det}, is important for the operations 
of pixelated tracking devices in a collider environment. 
%A value of $6.2 \times 10^{12}~1\,\text{MeV}~\text{n}_\text{eq}/\text{cm}^2$ for NIEL effects per year is 
%estimated for the first detector layer close to the vertex.  
%
The main effects on macroscopic sensor properties are\cite{pixel_det, RD48}
\begin{itemize}
\item[$\bullet$] Increase of the leakage current.
\item[$\bullet$] Increase of the biasing voltage on diode needed to achieve a given thickness of the active area. 
In other words, the depleted region gets thinner with the same biasing conditions. 
\item[$\bullet$] Charge trapping due to defects, which reduce the charge collection efficiency.
\end{itemize}
A value of $6.2 \times 10^{12}~1\,\text{MeV}~\text{n}_\text{eq}/\text{cm}^2$ for the NIEL  per year is 
estimated for the first layer of CEPC vertex detector\cite{cepcphys}.  
Thus,
the \supixi sensor has been irradiated under the fluence of $1 \times 10^{13}~1\,\text{MeV}~\text{n}_\text{eq}/\text{cm}^2$ 
at China Spallation Neutron Source (CSNS) to study the radiation-induced effects on sensor performance. 
%
%In this work, radiation-induced effects on sensors are studied via TCAD\cite{sentaurus} 
%simulations and measurements with radioactive source of \X{Fe}{55}. 
%
%The manuscript is organized as follows: 
%\sect{sim} demonstrates the TCAD simulation on charge collection. 
%
%The sensor is irradiated under the neutron fluence of $1 \times 10^{13}~1\,\text{MeV}~\text{n}_\text{eq}/\text{cm}^2$ 
%at China Spallation Neutron Source~(CSNS) and variations of sensor properties are characterized by \X{Fe}{55} in 
%The measured results of sensor properties before and after irradiation are discussed in \sect{rad:fe55}.  
%
The measurements of the sensor performance with the radioactive source of \X{Fe}{55} before and after irradiation 
are demonstrated in \sect{rad:fe55}.  The test results, taking A0 as an example, are discussed. 
The radiation-induced effects on charge collection are also studied via TCAD simulation, elaborated in \sect{sim}.
We conclude the radiation hardness of \supixi both from measurements and simulations in the last section.

%_____________________________________________________________________________________________________________________________
\section{Measurements with \X{Fe}{55}}
\label{sect:rad:fe55}

The sensor is tested with the radioactive source of \X{Fe}{55} before and after irradiation. The test system and method are 
demonstrated in \refcite{supix_test}.  

%\begin{figure}[htb!]
%\centering
%\subfigure[]{
%\includegraphics[width=0.6\textwidth,origin=c]{test_sys}
%\label{fig:sys:a}
%}
%\subfigure[]{
%\includegraphics[width=0.6\textwidth,origin=c]{working_flow_su1_rad}
%\label{fig:sys:b}
%}
%\label{fig:sys}

%\caption{The test system of \supixi for the radioactive source measurements: 
%(a) the test setup and (b) the working flow.
%}
%\end{figure}

%In order to investigate the sensor properties and the radiation effects, a single chip test system 
%shown in \fig{sys:a} is developed. It is consisted of the readout electronics and the data acquisition 
%system (SupixDAQ). The readout electronics involves a device under test (DUT) board, an analog to 
%digital conversion (ADC) board, a field programmable gate array (FPGA, Xilinx Kintex-7 KC705) board 
% and a personal computer~(PC). 
%
%As shown in \fig{sys:b}, the analog output is driven out of the sensor on DUT board when configured. 
%Then it is digitized by 16-bit ADCs (LTC2323-16) on ADC board and packaged on FPGA board. The 
%packaged raw data will be processed by correlated double sampling method (CDS) in SupixDAQ, in 
%which subtractions of two successive frames are recorded. 

%\subsection{The pixel-wise sensor properties}
%\label{sect:rad:fe55:calib}

\begin{figure}[htb!]
  \centering
    \includegraphics[width=1.\textwidth]{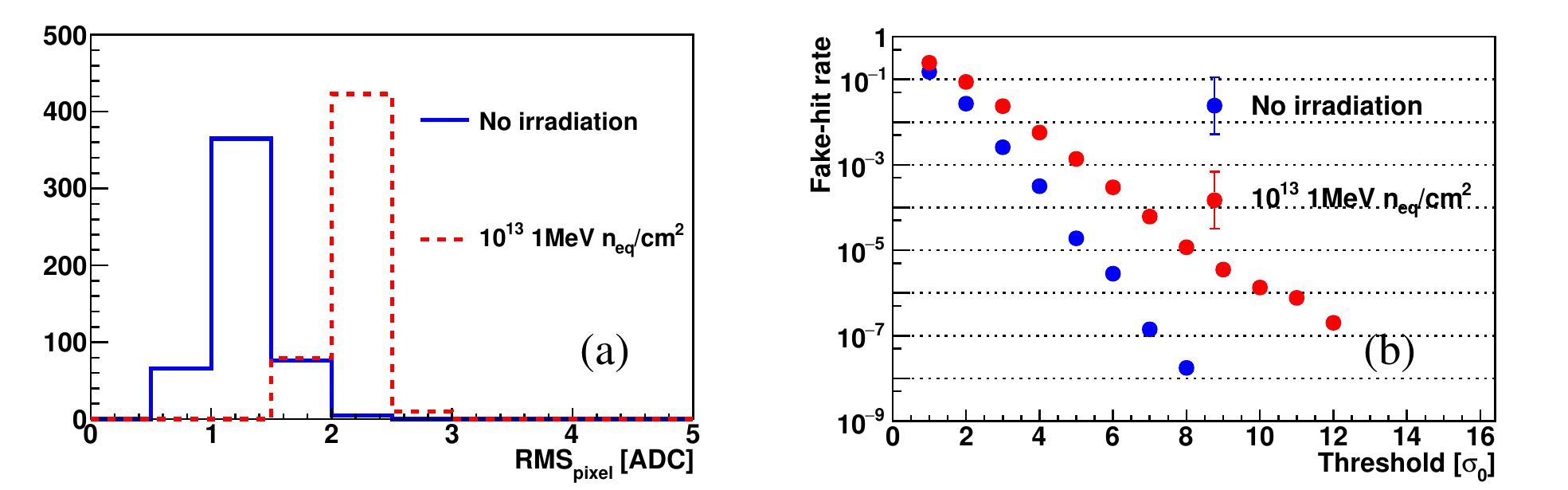}
    \caption{(a) Distributions of pixel-wise noise before 
    (blue solid line) and after (red dashed line) irradiation. 
    (b) The fake-hit rate versus the threshold before (blue bullet) 
    and after (red bullet) irradiation. $\sigma_0$ 
  is the RMS of the pixel output before irradiation.}
    \label{fig:noise:fakehit}
\end{figure}

Distributions of the fixed pattern noise before and after irradiation are shown in \fig{noise:fakehit}(a). 
The mean of the $\text{RMS}_\text{pixel}$ increases from 1.3 to 2.1 ADC counts (ADC) after irradiation. 
As a result, the fake-hit rate as a function of trigger threshold, demonstrated in \fig{noise:fakehit}(b), goes
up by one to three orders of magnitude after irradiation. 
%\fig{noise:fakehit}(b) demonstrates the fake-hit rate as a function of the threshold, where $\sigma_{0}$ is the 
%RMS of the pixel output before irradiation. The fake-hit rate can rise three orders of magnitude after irradiation.     

\subsection{The sensor gain calibration}
\label{sect:rad:fe55:calib}

The conversion gain (G) of the sensor is calibrated using the k-$\alpha$ (5.9 keV) soft 
X-ray of \X{Fe}{55} before and after irradiation. G is defined as
\begin{equation}
\text{G}~=~\frac{\text{ADC}}{\text{Q}},
\label{eq:gain}
\end{equation}
assuming 100\% of CCE when the X-ray photons absorbed in the depleted region of a single diode. Q is the 
charge generated. 1640 electron-hole pairs can be liberated when a k-$\alpha$ photon is absorbed in silicon, 
assuming that 3.6\,eV at room temperature is needed to create an electron-hole pair\cite{PDG}.

\begin{figure}[htb!] 
  \centering
  \subfigure[]{
    % \centering % \begin{center}/\end{center} takes some additional vertical space
    \includegraphics[width=0.47\textwidth,origin=c]{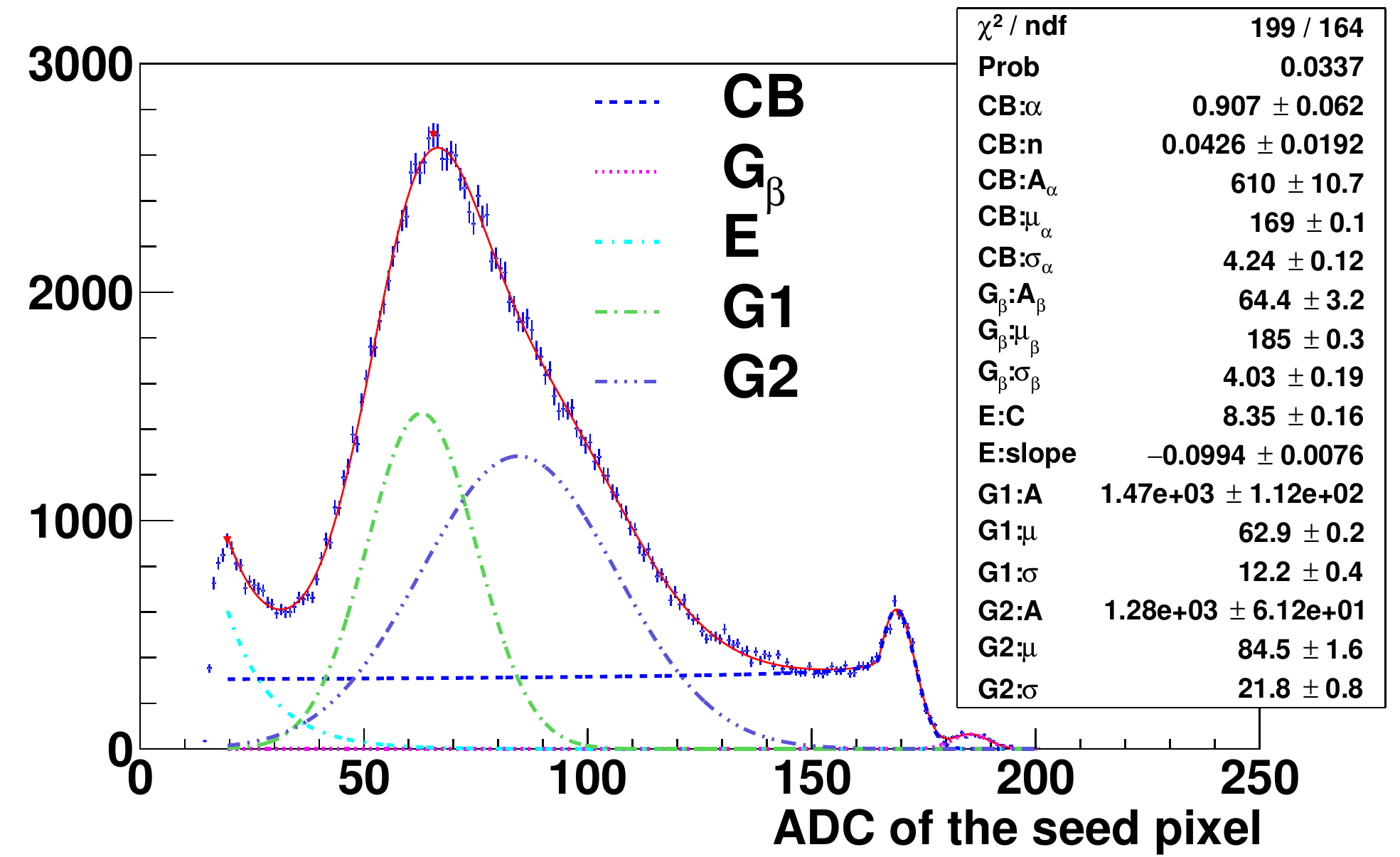}
    \label{fig:seedrad:a}
  }%
  \quad
  \subfigure[]{
    % \centering % \begin{center}/\end{center} takes some additional vertical space
    \includegraphics[width=0.47\textwidth,origin=c]{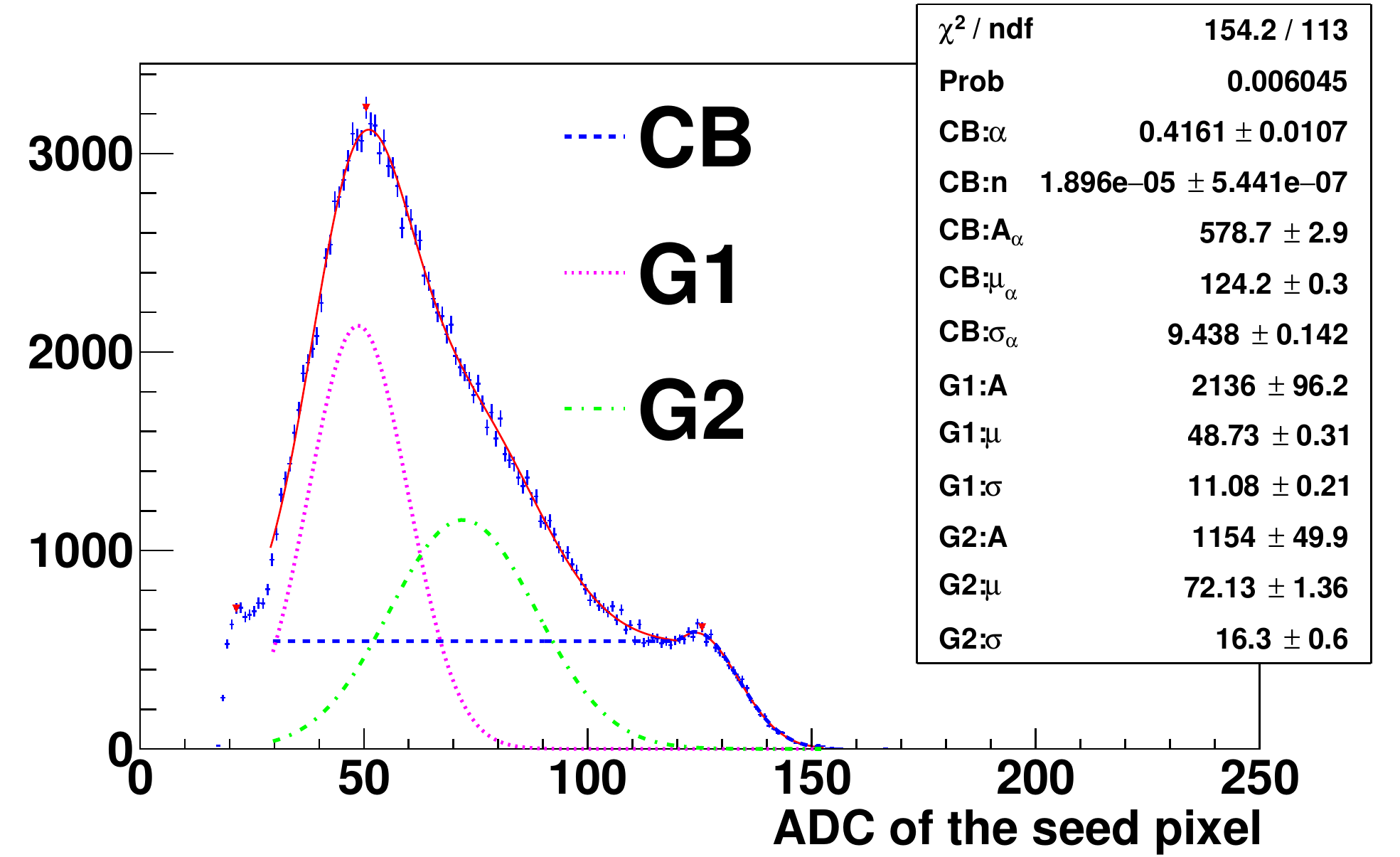}
    \label{fig:seedrad:b}
  }%
\label{fig:seedrad}
  \caption{Distributions of seed-pixel ADC (a) before and (b) after
   $1 \times 10^{13}~1\,\text{MeV}~\text{n}_\text{eq}/\text{cm}^2$ 
  neutron irradiation, together with full-spectrum fits.}
\end{figure}

The spectra of \X{Fe}{55}, reconstructed by seed pixels, 
before and after irradiation are shwon in \fig{seedrad:a} and \subref{fig:seedrad:b}, 
separately.    
Almost full spectra are well fitted by combinations of  component functions. At the right end of
the spectrum, the k-$\beta$ (6.5\,keV) peak disappears in \subref{fig:seedrad:b}. A single crystal ball function~(CB), 
instead of the combination of CB and Gaussian function ($\text{G}_{\beta}$) in \subref{fig:seedrad:a}, 
is applied to describe the full energy peak of k-$\alpha$. Peak with the highest amplitude (collection peak), 
contributed by the incident X-ray photons in the vicinity of the seed pixel, is well described by 
two Gaussian functions (G1 and G2) superposed with the tail of CB . The exponential function (E), describing 
the pedestal toward the left end of the spectrum in 
\subref{fig:seedrad:a}, fails in the fit in \subref{fig:seedrad:b}. 

The k-$\alpha$ peak shifts toward left by 
45 ADC, as a result of the increased diode capacitance after irradiation. The capacitance of a 
reversely biased
diode (C) shown in \fig{dev:side} can be estimated using the formula for parallel plate capacitor:
\begin{equation}
\text{C}~=~ \epsilon_{0}\epsilon_\text{{Si}}\frac{\text{A}}{\text{d}},
\label{eq:capa}
\end{equation}
where $\epsilon_0$ and $\epsilon_\text{Si}$ are vacuum permittivity and relative permittivity of silicon,
A and d are the area and depth of the depleted region. The radiation-induced decrease of d leads to 
the increase of C. Thus, the sensor output, Q/C, decreases.

\begin{table}[htb!]
  \caption{The geometrical parameters of \supixi sensor and the test results of 
  sensor gains and ENCs of respective sectors, 0 represents the
  sensor before irradiation while $10^{13}$ stands for the sensor 
  after $1 \times 10^{13}~1\,\text{MeV}~\text{n}_\text{eq}/\text{cm}^2$
  neutron irradiation.  }
  \label{tab:calib}
  \centering
  \begin{tabular}{c|c|c|c|c|c|c}
    \hline
    \multicolumn{2}{c|}{Sector}& A0 & A2 & A5  & A7 & A8\\
        \hline
        \multicolumn{2}{c|}{Surface ($\mu$m$^2$)} & 8 & 8 & 12 & 20 & 20\\
        \hline
        \multicolumn{2}{c|}{Footprint ($\mu$m$^2$)} & 11 & 11 & 18 & 44 & 50\\
        \hline\hline
         %\multirow{2}{*}{G} 
         G & 0 & 0.103$\pm$0.003 & 0.123$\pm$0.003 & 0.100$\pm$0.003 & 0.110$\pm$0.002 &0.112$\pm$0.003\\
         \cline{2-7}
         (ADC/e) & $10^{13}$ & 0.076$\pm$0.006 & 0.085$\pm$0.006 & 0.061$\pm$0.007 & 0.060$\pm$0.007 &0.049$\pm$0.007\\
        \hline
        %%\multirow{2}{*}{ENC} 
        ENC & 0 & 13$\pm$3 & 12$\pm$2 & 13$\pm$3 & 12$\pm$4 &12$\pm$2 \\
        \cline{2-7}
        (e) & $10^{13}$ & 28$\pm$3 & 27$\pm$3  & 42$\pm$6 & 50$\pm$6 &70$\pm$10\\
        \hline

  \end{tabular}

\end{table}

The conversion gains calibrated with k-$\alpha$ peaks for all sectors are listed in \tab{calib}. 
All sectors suffer considerable degradations after irradiation. The largest decrease of G, 56.3\%, 
is found in A8 which has the biggest diode dimension. Corresponding decreases for A2 and A5 are 30.9\% 
and 39.0\% respectively. A7 which has the same diode surface area as A8 does but smaller diode footprint 
area, suffers less decrease of 45.5\% than A8 does. 
Degradations on G show that bigger diode is more sensitive to the radiation-induced effects.
The difference of decrease between A0 and A2  is less than 5\%, indicating weak dependence 
on pixel pitch.

%\begin{table}[htb]
 % \centering
 % \label{tab:norad}
%  \caption{\label{tab:result:calib}Summaries of conversion gains and ENCs of sensors before and after 
 % irradiation.}
 % \smallskip
 % \begin{tabular}{|r|c|c|c|c|c|}
 %   \hline
  %  Sector& \multicolumn{2}{c}{No irradiation} \vline & \multicolumn{2}{c}{$10^{13}~$1\,\text{MeV}
 %   $\text{n}_\text{eq}/$\text{cm}$^2$} \vline  \\
   % \hline
  %  & G~(ADC/$\text{e}^-$) & ENC~($\text{e}^-$)  & G~(ADC/$\text{e}^-$) & ENC~($\text{e}^-$) \\
  %  \hline
   % A0 & $0.103 \pm 0.003$ & $13 \pm 2.6$ & $0.076 \pm 0.006$ & $28.1 \pm 2.9$\\
   % A2 & $0.123 \pm 0.003$ & $12 \pm 2.4$ & $0.085 \pm 0.006$ & $27.2 \pm 2.6$\\
   % A5 & $0.100 \pm 0.003$ & $13 \pm 2.8$ & $0.061 \pm 0.007$ & $41.8 \pm 5.5$\\
%    A7 & $14 \pm 0.4$ & $12 \pm 2.4$ & $8.1 \pm 0.9$ & $47 \pm 2.0$\\
   % A8 & $0.112 \pm 0.003$ & $12 \pm 2.4$ & $0.049 \pm 0.007$ & $69.7 \pm 10.3$\\   
     % \hline
  %\end{tabular}
%\end{table}
The fixed pattern noise in \fig{noise:fakehit}(a) is transformed into ENC after G being calibrated.
%\begin{equation}
%\text{ENC} = \frac{\mean{\text{RMS}_\text{noise}}}{\text{G}},
%\label{eq:enc}
%\end{equation}
%where $\mean{\text{RMS}_\text{noise}}$ is the mean of $\mean{\text{RMS}_\text{CDS}}$ in \fig{noise:fakehit}(a).
As listed in \tab{calib}, for small diodes, increases of ENCs for A0 and A2 are 115\% and 125\% respectively after irradiation. 
For medium size diodes, A5 increases by 223\%.  
A7 and A8, which has the largest diode surface area, suffer the increase of ENC over 300\%. 
And the increase of A8 is 50\% more than that of A7 due 
to the larger area of footprint.

\begin{figure}[htb!]
  \centering
    \includegraphics[width=.7\textwidth]{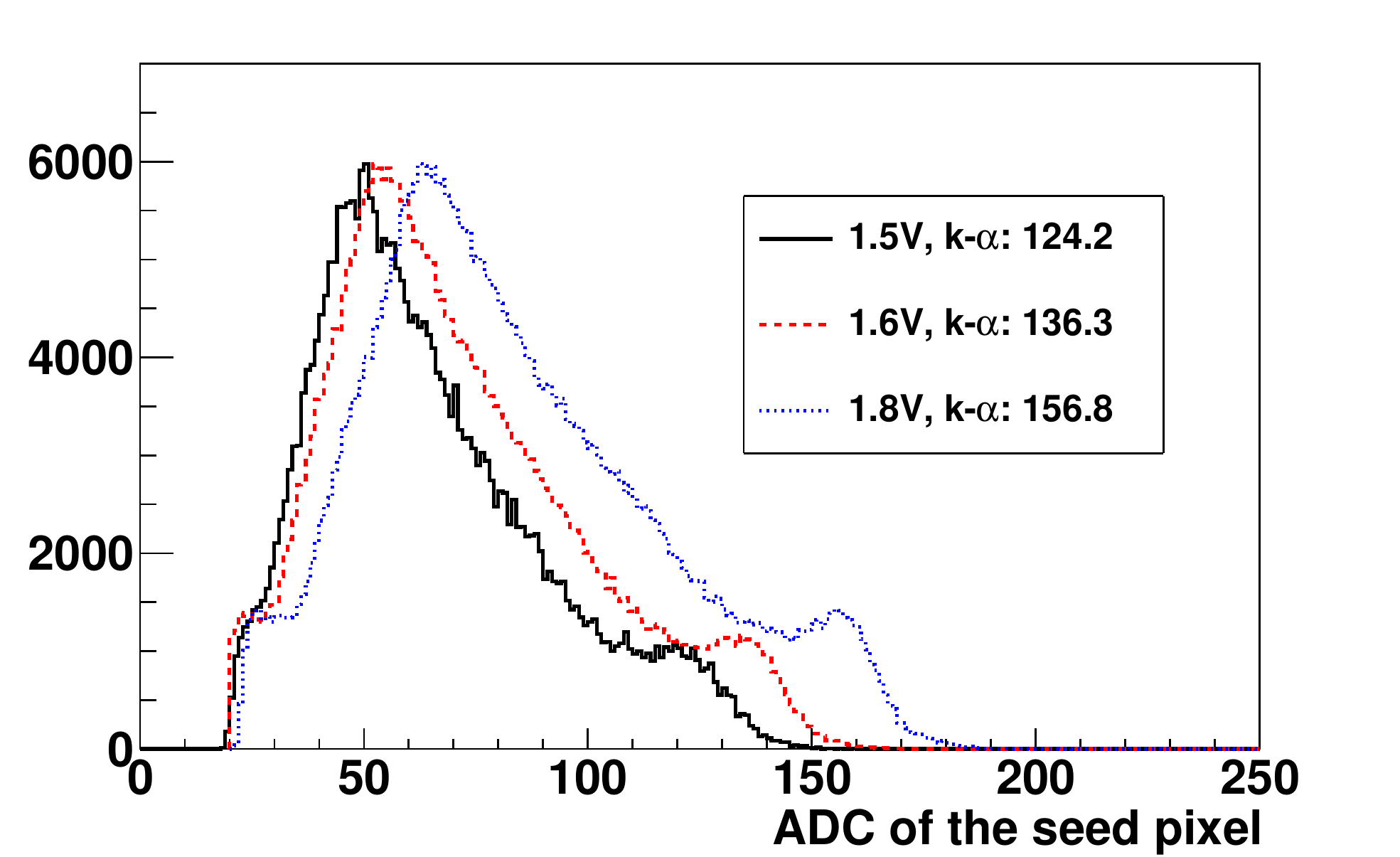}
    \caption{Distributions of seed-pixel ADC under various biased voltages.}
    \label{fig:calib:multi}
\end{figure} 

The pixel-wise CCE SNR are defined in \refcite{supix_test}.
Test results for all sectors before and after irradiation are summarized in \tab{result}. CCEs, for the same sector, 
are consistent within uncertainties before and after irradiation. However, SNRs suffer decreases after irradiation. 
Decreases on SNR exhibit strong dependence on diode geometry. Bigger the diode is, larger gets the ENC after irradiation.
Consequently, larger diodes suffer more decreases on SNR.

%A8, equipped with the largest diodes, suffers the increase of the ENC over four times. 

%The pixel-wise CCE and SNR are evaluated using the fitted location of the collection peak 
%\begin{subequations}
%\begin{equation}
%\text{CCE}_\text{pixel} = \frac{\text{ADC}_\text{MPV, seed}}{\text{ADC}_{\text{k-}\alpha}},
%\label{eq:pixcce}
%\end{equation}
%\begin{equation}
%\text{SNR}_\text{pixel} = \frac{\text{ADC}_\text{MPV, seed}}{\mean{\text{RMS}_\text{noise}}},
%\label{eq:pixsnr}
%\end{equation}
%\end{subequations}
%where $\text{ADC}_{\text{k-}\alpha}$ is the location of the k-$\alpha$ peak. 
%The pixel-wise charge collection efficiency (CCE) is evaluated using the fitted locations of the collection peak and k-$\alpha$ peak. 
%Results for all sectors are summarized 
%in \tab{result}. The irradiated and the non-irradiated sensors are consistent within the uncertainties on 
%$\text{CCE}_\text{pixel}$, indicating the negligible impacts on pixel-wise charge collection of \supixi sensor 
%under the fluence of $10^{13}~1\,\text{MeV}~\text{n}_\text{eq}/\text{cm}^2$. However, $\text{SNRs}_\text{pixel}$ 
%suffer great decreases, over 50\%, as results of the increased ENCs. Consistent with the ENCs, large diodes 
%are more sensitive to the radiation.   

We have increased the biased voltage from 1.5\,V (default) to 1.8\,V (maximum allowed) to see the difference 
of sensor performance. The k-$\alpha$ peak, A0 as an example, has a right-shift with the voltage going up, 
so as the collection peak, as shown in \fig{calib:multi}. Correspondingly, the sensor gain increases from 0.076\,ADC/e 
(1.5\,V) to 0.096\,ADC/e (1.8\,V) and the ENC decreases from 28\,$\text{e}^-$ to 21\,$\text{e}^-$. 

\subsection{The cluster-wise sensor properties}
\label{sect:rad:fe55:cluster}

\begin{figure}[htb!]
  \centering
    \includegraphics[width=.98\textwidth]{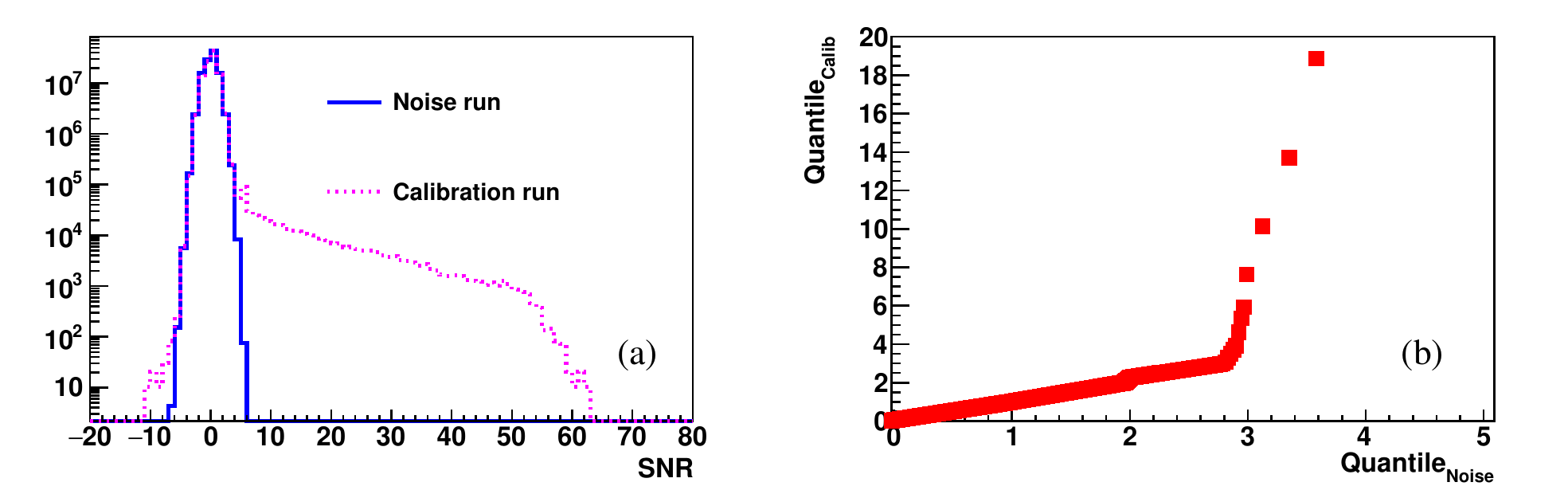}
    \caption{The determination of the clustering threshold after 
    neutron irradiation: distributions of the pixel SNR with 
    (calibration run) and without (noise run) \X{Fe}{55}, (b) 
    quantiles of the pixel-wise SNR distribution in calibration run 
    as a function of that in noise run. }
    \label{fig:cluster:quantile}
\end{figure}

Mostly, charge carriers ionized by incident photons of \X{Fe}{55} are collected by serval pixels through 
thermal diffusions in the epitaxial layer, forming clusters. The cluster properties and the radiation-induced 
effects are studied with a self developed reconstruction algorithm introduced in \refcite{supix_test}.   
In \refcite{supix_test}, the clustering threshold for fired pixels, based on pixel SNR, 
is set to be 3.1 for less than one per mil of noise pixels leaked into the cluster. A new determination of the 
clustering threshold, shown in \fig{cluster:quantile}, is developed in this work. Distributions of pixel 
SNR with (calibration run) and without (noise run) \X{Fe}{55} are shown in (a). Quantiles of the pixel SNR
distribution in calibration run as a function of that in noise run are shown in (b). The function is not linear 
with slop of 1 any more when SNR > 2.8,  , which means the signal for X-ray of \X{Fe}{55} dominates. Thus, SNR of 2.8 is 
selected as the clustering threshold in this paper. 

%\begin{figure}[htb!] 
 % \centering
    % \centering % \begin{center}/\end{center} takes some additional vertical space
 %   \includegraphics[width=1.\textwidth,origin=c]{cluster_info_cmp}
 %   \label{fig:clusterinfo}
 % \caption{Variations of the cluster properties: (a) the cluster size and (b) the cluster 
  %SNR before (blue) and after (rad) irradiation, and the ADC sum  of the cluster 
 % before (c) and after (d) irradiation.  
 % }
%\end{figure}

%The cluster reconstruction starts with a seed pixel. Neighboring pixels fired, diagonal ones included, 
%are selected. The selection repeats after the extrapolation center shifting to selected neighbors 
%whose output are zeroed until there are no neighbors fired anymore or reaching the edge of the pixel array. 
%
%The clustering threshold for fired pixels, based on SNR of the selected pixel, is set to be 3.1 for less than 
%one per mil of noise pixels leaked into the cluster. 
%

\begin{figure}[htb!]
  \centering
    \includegraphics[width=.9\textwidth]{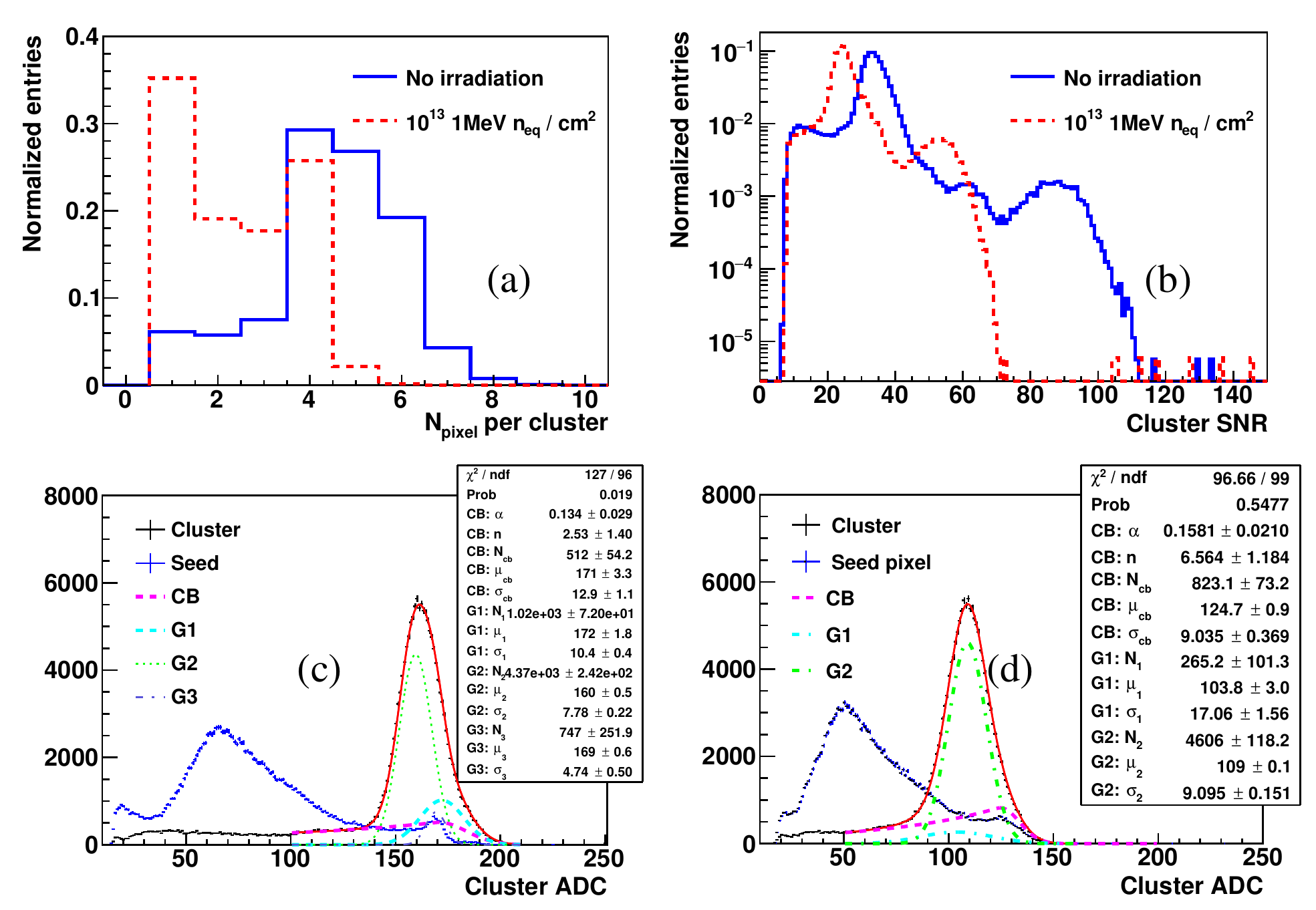}
    \caption{Variations of the cluster properties: (a) the cluster size and 
    (b) the cluster SNR before (blue) and after (rad) irradiation, and 
    the ADC sum  of the cluster before (c) and after (d) irradiation.}
    \label{fig:cluster:info}
\end{figure}

\begin{table}[htb!]
  \caption{Geometrical parameters of the \supixi sensor and the 
  test results with \X{Fe}{55} radioactive source, 0 represents the
  sensor before irradiation while $10^{13}$ stands for the sensor 
  after the $1 \times 10^{13}~1\,\text{MeV}~\text{n}_\text{eq}/\text{cm}^2$
  of the neutron irradiation.  }
  \label{tab:result}
  \centering
   \begin{tabular}{c|c|c|c|c|c|c|c}
    \hline
    \multicolumn{3}{c|}{Sector}& A0 & A2 & A5  & A7 & A8\\
        \hline\hline
        \multicolumn{3}{c|}{Sensitive Area x (mm)} & 0.3 & 1.3 & 1.3  & 0.6 & 1.3\\
        \hline
         \multicolumn{3}{c|}{Sensitive Area y (mm)} & \multicolumn{5}{c}{1.3} \\
        \hline
        \multicolumn{3}{c|}{Pixel Pitch x ($\mu$m)} & 21 & 84 & 84  & 42 & 84 \\
        \hline
        \multicolumn{3}{c|}{Pixel Pitch y ($\mu$m)} & \multicolumn{5}{c}{21} \\
        \hline
        \multicolumn{3}{c|}{Diode Surface ($\mu$m$^2$)} & 8 & 8 & 12  & 20 & 20  \\
        \hline
        \multicolumn{3}{c|}{Diode Footprint ($\mu$m$^2$)} & 11 & 11 & 18 & 44 & 50\\
        \hline\hline
%       	\multirow{8}{*}{Pixel} & \multirow{4}{*}{0} & G & 0.103$\pm$0.003 & 0.123$\pm$0.003 & 0.100$\pm$0.003 & 0.112$\pm$0.003  \\
%	\cline{3-8}
%						&				& ENC & 13$\pm$3 & 12$\pm$2 & 13$\pm$3 & 12$\pm$2 \\
%	\cline{3-8}
	\multirow{4}{*}{Pixel} & \multirow{2}{*}{0} & CCE& 39.1$\pm$1.0 & 31.7$\pm$0.7 & 27.5$\pm$0.6 & 30.4$\pm$0.8 & 37.2$\pm$0.7\\
	\cline{3-8}
						&				& SNR & 52$\pm$11 & 49$\pm$10 & 36$\pm$8 & 42$\pm$9 & 53$\pm$11\\
	\cline{2-8}
%						& \multirow{4}{*}{$10^{13}$} & G & 0.076$\pm$0.006 & 0.085$\pm$0.006 & 0.061$\pm$0.007 & 0.049$\pm$0.007\\
%	\cline{3-8}
%						&						& ENC & 28$\pm$3 & 27$\pm$3  & 42$\pm$6 & 70$\pm$10\\
	\cline{3-8}
						&\multirow{2}{*}{$10^{13}$}& CCE & 40.1$\pm$1.5 & 31.8$\pm$1.0 & 26.0$\pm$1.1 & 31.6$\pm$1.1 & 37.7$\pm$1.2 \\
	\cline{3-8}
						&						& SNR & 23$\pm$2 & 19$\pm$1 & 10$\pm$1 & 10$\pm$1 &9$\pm$1\\
	\hline
	\multirow{4}{*}{Cluster} & \multirow{2}{*}{0} & CCE & 95.6$\pm$0.5 & 90.3$\pm$0.4 & 93.8$\pm$0.5 & 96.8$\pm$0.4 &92.4$\pm$0.6\\
	\cline{3-8}
						&				& SNR & 36$\pm$3 & 33$\pm$2 & 32$\pm$2 & 36$\pm$3 & 36$\pm$3\\
	\cline{2-8}
						& \multirow{2}{*}{$10^{13}$} & CCE & 89.0$\pm$1.0 & 75.2$\pm$0.9 & 73.9$\pm$1.5 & 90.6$\pm$1.7 &73.2$\pm$1.7\\
	\cline{3-8}
						&						& SNR & 22$\pm$2 & 19$\pm$2 & 12$\pm$2 & 14$\pm$2 & 10$\pm$2\\
       	\hline
  \end{tabular}

\end{table}

The cluster properties before and after irradiation are demonstrated in \fig{cluster:info}. 
The mean of the cluster size, shown in (a), decreases from 4.4 to 2.4 after irradiation, 
as a result of charge trapping. 
%
%The cluster SNR, shown in (b), is defined as 
%\begin{equation}
%\text{SNR}_\text{cluster} = \frac{\sum_i \text{ADC}_i}{\sqrt{\sum_i \sigma_i^2}},
%\label{eq:clustersnr}
%\end{equation}
%where $\text{ADC}_i$ and $\sigma_i$ are the ADC and the noise of the $i$-th pixel.
%the sum of the pixel output over square root of the quadratic sum of  the pixel noise in cluster, 
%also shows a decrease in \fig{clusterrad:c}.
%
The cluster SNR, shown in (b), also decreases, but most of clusters have the SNR over 20. 
Distributions of the cluster ADC before and after irradiation are shown in (c) and (d) respectively, 
together with the distribution of seed-pixel ADC for comparison. The distributions are well fitted 
by Gaussian functions superposed a crystal ball function. 
%The cluster CCE is defined as
%\begin{equation}
%\text{CCE}_\text{cluster} = \frac{\text{ADC}_\text{MPV, cluster}}{\text{ADC}_{\text{k-}\alpha}},
%\label{eq:cluster:charge}
%\end{equation}
%where $\text{ADC}_\text{MPV, cluster}$ is the location of the cluster ADC peak.  

The cluster-wise CCE and SNR, defined in \refcite{supix_test}, are measured 
for the irradiated sensor. Results along with that before irradiation are summarized in \tab{result}. 
Before irradiation, the cluster properties show weak dependence on geometrical configuration. On the one hand, 
the SNRs are consistent within uncertainties. On the other hand, variations of CCEs are within about 5\% referring 
to the CCE of A0. 
The decreases of CCE after irradiation show the significance of pixel pitch in charge collection.
 The decrease of CCE for A0 is less than 7\% but 17\% for A2. Variations of decreases among sectors with largest x-pitch 
 are less than 5\%.  
 A7 has larger diode dimensions than A0 does and the pixel pitch of A7 is twice of A0's. 
 A7 gets the similar decrease on CCE as A0 does.
 %A7 gets the smallest decrease of CCE, 
 %implying that large diode dimensions help on charge collection after irradiation with a constrained pixel pitch 
 %(42\,$\mu$m in this work).   
 The decreases of SNR are apparent after irradiation for all sectors. 
 The decreases of pixel-wise and cluster-wise SNR exhibit similar dependence on diode geometry.
% Bigger the diode is, larger the ENC after irradiation is. Consequently, the SNR suffers greater loss in sectors with larger diode.    
 
 %
 % SNRs vary a lot because of the significant variations of ENCs after irradiation, and A8 performs the worst. 

\section{TCAD simulations}
\label{sect:sim}
The radiation-induced damage, under fluence of $1\times10^{13}~1\,\text{MeV}~\text{n}_\text{eq}/\text{cm}^{2}$, on 
charge collection are studied via TCAD. 
Simulations are consisted of device construction and emulations of electric properties of devices with 
physics models. 

\begin{figure}[htb!] 
	\centering
	\subfigure[]{	
	\includegraphics[width=.27\textwidth,origin=c]{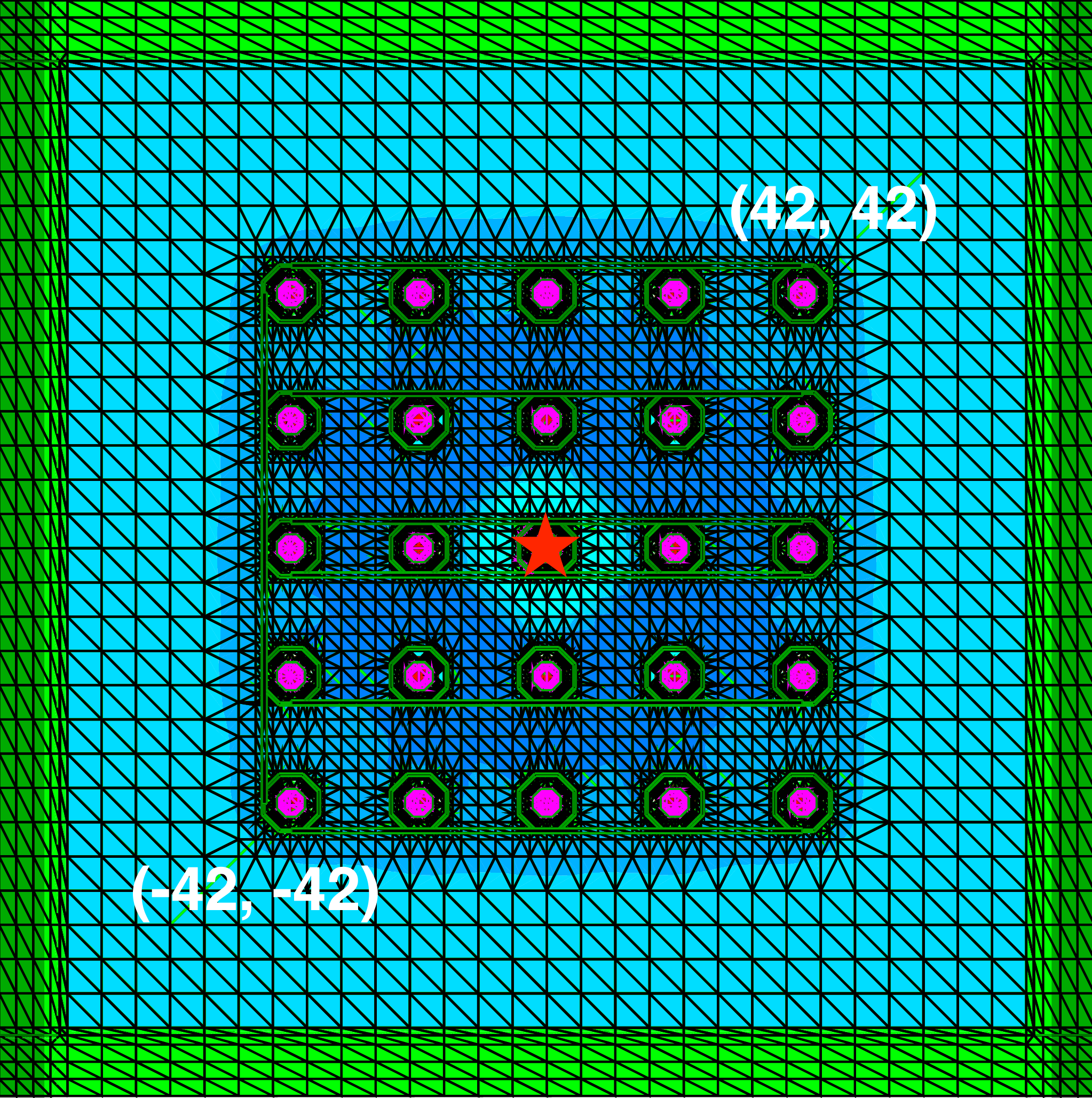}
	% "\includegraphics" from the "graphicx" permits to crop (trim+clip)
	% and rotate (angle) and image (and much more)
	\label{fig:dev:matrix}
	}
	\subfigure[]{	
	\includegraphics[width=.55\textwidth,origin=c]{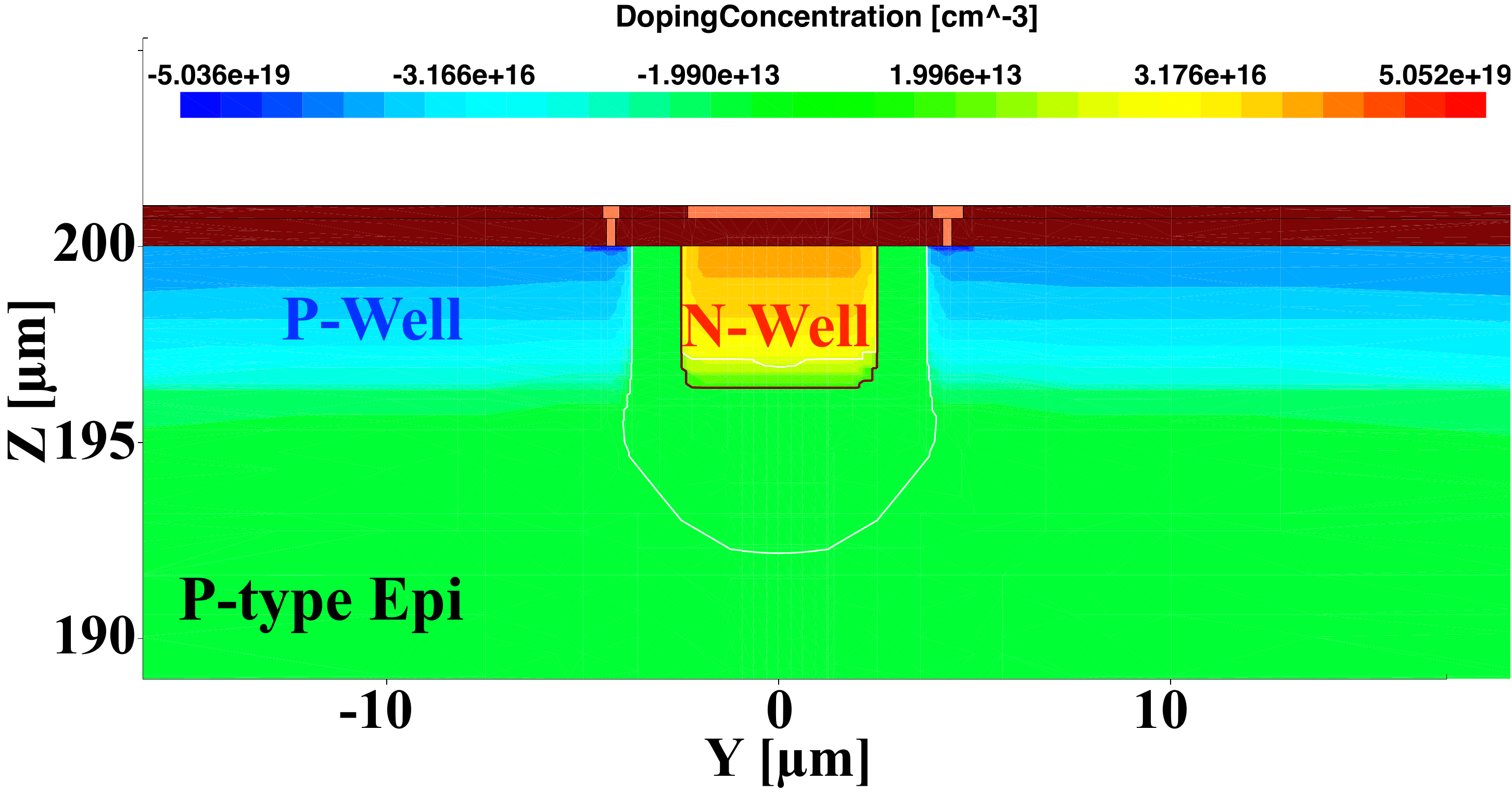}
	% "\includegraphics" from the "graphicx" permits to crop (trim+clip)
	% and rotate (angle) and image (and much more)
	\label{fig:dev:side}
	}

	\caption{(a) the topview of a simulated device with $5 \times 5$ pixel array 
	in A0 and (b) the doping profile of a single diode in Y-Z 
	plane.}
\end{figure}

The device construction consists of geometry definition and doping profile. $5 \times 5$ pixel matrixes with the same
diode geometries and pixel arrangements in sectors measured are constructed. 
As an example, the top view of a $5\times5$ pixel matrix in A0 is shown in \fig{dev:matrix}, in which the X-Y co-ordinate of the 
diode at bottom-left is (-42, -42) in the unit of $\mu$m, as a reference. 
The device is divided into small building blocks by 
doping concentration dependent meshing strategy, shown as black lines, for finite element analyses. With the limits of 
the CPU~(Intel Core i7-6700 \@ 3.4\,GHz) and the random access memory~(48\,GB~DDR4), total vertices and elements of 
building blocks are controlled within 200\,000 and 1\,000\,000 respectively. Additional volumes of silicon~(green) with 
low doping concentration~($< 10^5~\text{cm}^{-3}$) are added to boundaries due to the reflective boundary 
conditions\cite{boundary_condition}. 

\fig{dev:side} shows the doping concentration of a single diode in Y-Z plane. The doping concentration of phosphorus 
(boron) is higher, as the color of the doped area gets redder (bluer). The doping strategy of N-well~(P-well) along Z axis 
is a combination of multiple Gaussian functions with peak concentrations ranging from $4.5 \times 10^{17} \text{cm}^{-3}$ 
to $5 \times 10^{19} \text{cm}^{-3}$. A reversely biased voltage of 1.5\,V is applied on the N-well diode, forming the 
 depleted region with the depth of 4.9\,$\mu$m, enclosed by the white lines in \fig{dev:side}. 
 The thickness of the device is 200\,$\mu$m.

%Reasonable simulations are results of proper physics models. Charge generation and 
%radiation models are essential in this work. 
A heavy ion model 
is applied to simulate the charge generation of minimum ionizing particles~(MIPs) in
silicon. 
In the model, the linear energy transfer parameter is set to be 
$1.28 \times 10^{-5}~\text{pC}/ \mu$m for the estimated most probable energy 
loss rate of 80 electron-hole pairs$/\mu$m for MIPs in 200\,$\mu$m silicon\cite{PDG}. 
Simulation vary with the impinging positions of incident particles. Two typical 
cases are chosen for discussion. The first one is (0, 0) in X-Y plane~(case 1), shown as 
the red star in \fig{dev:matrix}, on which a pixel diode is hit. The other one is complicated. 
Positions in case 2 for A0 and A7 are (10, 10) and (21, 10) respectively. They locate the center
of four pixels. For sectors with x-pitch of 84\,$\mu$m, (-25, 0) which is the center of 3 pixels, 
shown in \fig{chip:c}, is selected.   
%(10, 10) ((25, 0) for A2, A5 and A8, shown in \fig{chip:c}) which is the center of 4 pixels~(3 for 
%A2, A5 and A8)~(case 2). 
%
The incident direction is along Z axis and the length of the track is 200\,$\mu$m.

For bulk damage on the sensor, a three level radiation damage model is applied\cite{3level_model}. In this model, 
radiation induced traps are distributed on different energy levels, 
as listed in \tab{radiation:model}. $\text{E}_\text{c}$ is the energy level of conduction band of 
silicon while $\text{E}_\text{v}$ is for the valence band. $\sigma_\text{n}$~($\sigma_\text{p}$) 
is the cross-section between a trap and an electron~(hole) and $\eta$ is the introduction rate. 
The concentration of radiation induced traps ($\text{N}_\text{traps}$) is given by
\begin{equation}
\text{N}_\text{traps}~=~\eta \Phi_\text{eq},
\label{eq:model:trapconc}
\end{equation}
where $\Phi_\text{eq}$ is the neutron equivalent fluence.

\begin{table}[htbp]
	\centering
	\caption{\label{tab:radiation:model}The parameters of three-level radiation damage model. 
	$\text{E}_\text{c}$ represents the energy level of conduction band while $\text{E}_\text{v}$ is 
	for the valence band. $\sigma_\text{n}$~($\sigma_\text{p}$) is the cross-section between a trap 
	and an electrons~(hole) and $\eta$ is the introduction rate.}
	\smallskip
	\begin{tabular}{rccc}
		\hline
		Energy level & $\sigma_\text{n}$($\text{cm}^{-2}$) & $\sigma_\text{p}$($\text{cm}^{-2}$) & 
		$\eta$($\text{cm}^{-1}$)\\
		\hline
		$\text{E}_\text{c}\,-\,0.42~\text{eV}$& 2.0$\times$$10^{-15}$ &2.0$\times$$10^{-14}$ & 1.613\\
		$\text{E}_\text{c}\,-\,0.46~\text{eV}$&5.0$\times$$10^{-15}$& 5.0$\times$$10^{-14}$& 0.9\\
	    	$\text{E}_\text{v}\,+\,0.42~\text{eV}$& 2.5$\times$$10^{-14}$ &2.5$\times$$10^{-15}$ & 0.9\\
		\hline
	\end{tabular}
\end{table}

\begin{figure}[htb!]
  \centering
    \includegraphics[width=.7\textwidth]{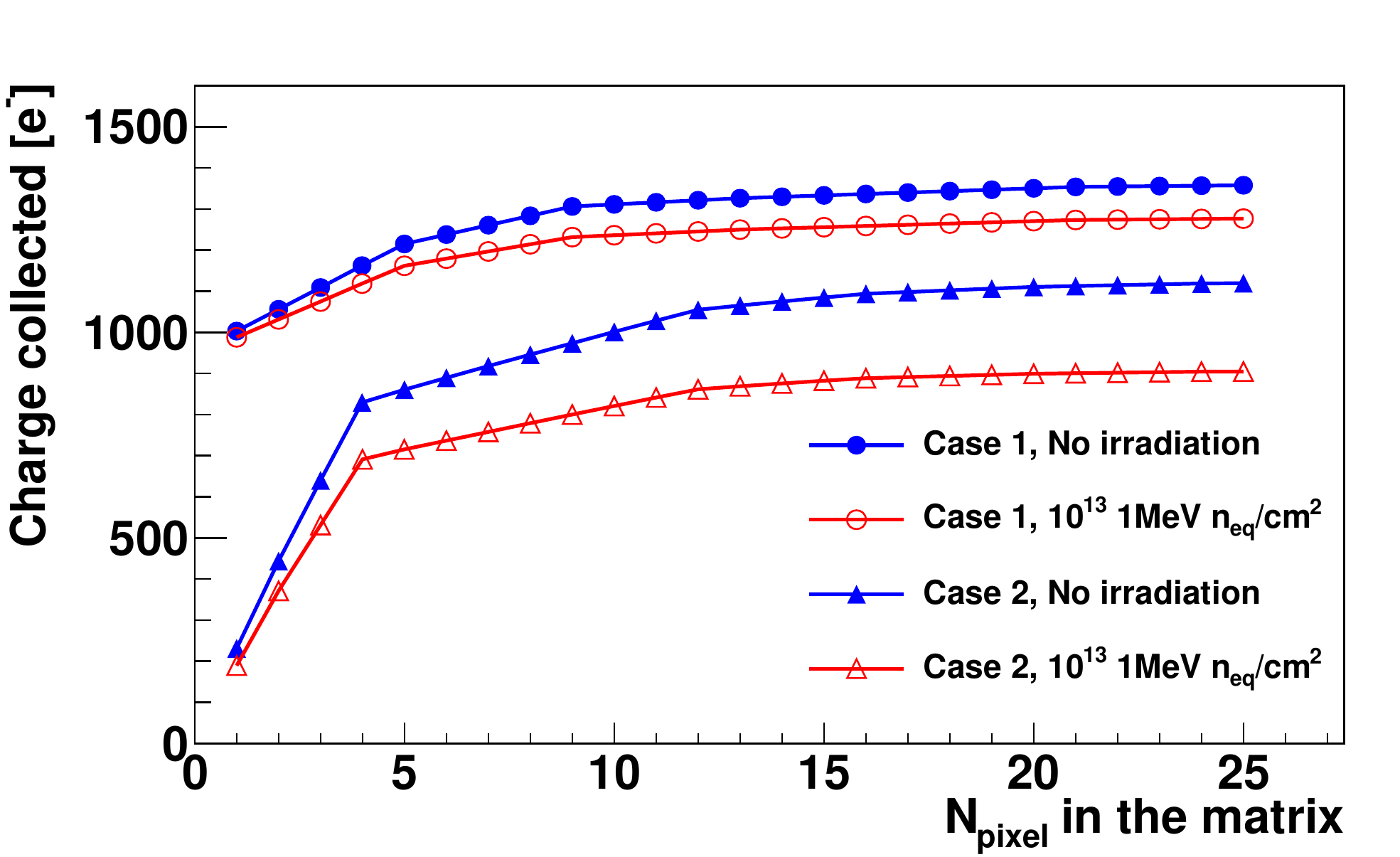}%%charge_sim_combined}
    \caption{Simulated charge collections as functions of pixels in the matrix 
    before (blue) and after (rad) irradiation. Case 1 is represented by the marker of circle 
    while case 2 is shown as triangle.}
    \label{fig:charge:sim}
\end{figure}

%\begin{table}[htb]
 % \centering
 % \label{tab:rad}
 % \caption{\label{tab:simresult:loss} Decreases of simulated charge collections. }
 % \smallskip
 % \begin{tabular}{|r|c|c|c|c|c|c|c|c|c|c|c|}
   % \hline
   %  Sector &    \multicolumn{4}{c}{Decrease of charge collected after irradiation~(\%)}\vline  \\
   % 	\hline
%	& \multicolumn{2}{c}{Hit on central pixel diode} \vline& \multicolumn{2}{c}{Hit on center of 4~(3) pixels}\vline \\   
%	\hline
%	&Seed pixel & Matrix & Seed pixel & Matrix\\
%	\hline
%	A0 & 1.5 & 6.0 & 18.2 & 19.2\\
%	A2 & 6.9 & 12.2 & 64.5 & 68.7\\
%	A5 &  3.5 & 5.6 & 58.5 & 60.2 \\
%	A8 & 4.4 & 8.2 & 59.0 &  65.2\\
%	\hline
 % \end{tabular}
%\end{table}

\begin{table}[htb!]
  \caption{Geometrical parameters of \supixi sensor, and the 
  charge collection simulation before and after irradiation. See 
  the text for explanation.} 
  \label{tab:simresult:loss}
  \centering
    \begin{tabular}{c|c|c|c|c|c|c|c|c}
    \hline
    \multicolumn{4}{c|}{Sector}& A0 & A2 & A5 & A7 & A8\\
        \hline\hline
        \multicolumn{4}{c|}{Pixel Pitch x ($\mu$m)} & 21 & 84 & 84 & 42 & 84 \\
        \hline
        \multicolumn{4}{c|}{Pixel Pitch y ($\mu$m)} & \multicolumn{5}{c}{21} \\
        \hline
        \multicolumn{4}{c|}{Diode Surface ($\mu$m$^2$)} & 8 & 8 & 12  & 20 & 20  \\
        \hline
        \multicolumn{4}{c|}{Diode Footprint ($\mu$m$^2$)} & 11 & 11 & 18  & 44 & 50\\
        \hline\hline
        \multirow{8}{*}{Q (e)} & \multirow{4}{*}{0} & \multirow{2}{*}{Case1} & Seed & 1003 & 1117 & 1231 & 1235 & 1287\\
        	\cline{4-9}
        		&				&	& Matrix & 1358 & 1227 & 1321 & 1467 & 1463 \\	%231 & 155 & 188 & 256 \\
	\cline{3-9}
		&				& \multirow{2}{*}{Case2} & Seed & 231 & 155 & 188 & 179 & 256 \\	%1358 & 1227 & 1321 & 1463 \\
	\cline{4-9}
		&		(1\,MeV $\text{n}_\text{eq}$/cm$^2$)	&		& Matrix & 1120 & 521 & 636 & 821 &748\\
	\cline{2-9}
	
	& \multirow{4}{*}{$10^{13}$} & \multirow{2}{*}{Case1} & Seed & 988 & 1040 & 1188 & 1201 & 1226\\
	\cline{4-9}
        		&									&	& Matrix & 1276 & 1077 & 1202 & 1367 & 1303\\		%189 & 55 & 78 & 105\\
	\cline{3-9}
		&			& \multirow{2}{*}{Case2} & Seed & 	189 & 55 & 78 & 92 & 105\\		%1276 & 1077 & 1246 & 1343\\
	\cline{4-9}
		&	(1\,MeV $\text{n}_\text{eq}$/cm$^2$)	&		& Matrix & 905 & 163 & 266 & 398 & 260\\  
						  
	\hline
  \end{tabular}

\end{table}

Simulated charge collections, as functions of pixels after descending orders of electrons collected, 
are presented in \fig{charge:sim}. Charge collected for seed pixels ($\text{N}_\text{pixel} = 1$) 
are consistent before and after irradiation. For pixel matrix,  case 2 has a large decrease of Charge collected 
than case 1 does. 90\% of the charge collection are done within 5 pixels in case 1 and 10 in case2.
Results for all sectors are summarized in \tab{simresult:loss}.
In case 1, Decreases of charge collected for seed pixels are within 7\% after irradiation. 
Decreases for pixel matrixes are within 13\%.  
 A0 which has the smallest pixel pitch gets the smallest decrease, 6\%. 
 The decrease of A7 which has the medium size pixel pitch is 7\%, close to that of A0.
 Variations of decreases for large pixel sectors with various diode geometries are within 3\%.
In case 2, all sectors suffer considerable
loss on charge collection, over 48\%, both for seed pixel and matrix except A0.    
%A2 and A0 have the same diode geometry but the x-pitch of A2 is four times that of A0. 
%In case 1, the radiation caused decreases on charge collection for A2 are 7\% and 12\% for seed pixel 
%and matrix respectively while the corresponding values for A0 are 2\% and 6\%. In case 2, the collected charge 
%decreases over 60\% for both seed pixel and matrix in A2, wheras the decreases are below 20\% in A0. 
%A2 suffers greater decreases after irradiation both in seed pixel (pixel with largest output) and matrix compared to 
%A0, especially in case 2.  
%Variations of decreases among sectors with x-pitch of 84\,$\mu$m are within 9\% for 
%both seed pixel and matrix and both two impinging cases.  
%
The simulation indicates that diffusion length is the key point for the radiation induced signal loss. 
Longer the charge carriers diffuse, more possibly are they trapped by defects. Thus, small pixel pitch 
gains an edge on charge collection after irradiation. The test results are between typical cases simulated
and are close to case1.

%The diode geometry has less impacts on radiation induced degradations than the pixel pitch does. 

%In sectors with large pixel configuration, enlarged diodes help to reduce the radiation caused decrease of charge collection and
%A5 achieves the lowest decrease. However, variations of the decreases among large pixel sectors are less than that between A0 
%and A2, indicating less dependences on diode size for radiation induced impact on charge collection.
%Variations of the decreases of charge collections for different sectors with large pixel configurations are within 10\% under
%all circumstances. 

\section{Conclusion}
\label{sect:conclusion}
The radiation induced effects on \supixi, a CMOS pixel sensor, under the fluence of 
$1 \times 10^{13}~1\,\text{MeV}~\text{n}_\text{eq}/\text{cm}^2$ are studied via by 
measurements with radioactive source of \X{Fe}{55} and TCAD simulations. 
 In measurements, the sensor gain is calibrated using the k-$\alpha$ X-ray of \X{Fe}{55}. Then, the ENC and the pixel-wise
 CCE and SNR are obtained. The radiation-induced effects on properties of pixel clusters are studied with a preliminary clustering 
 algorithm.  A new determination of clustering threshold, based on the quantiles of the pixel-SNR distributions in noise run and 
 calibration run, is developed. 
 In simulations, two typical impinging positions of incident particles are selected and the corresponding charge collections
 in $5\times5$ pixel matrixes before and after irradiation are simulated.   

 The calibration of the sensor gain shows strong dependence on diode geometries. The larger diodes suffer more decreases of 
 sensor gain after irradiation. Degradations on ENC show the similar dependence, so do the pixel-wise and cluster-wise SNR 
 which are ENC related. 
 
 For radiation-induced effects on charge collection, the test results are between the typical cases in simulation, and are close 
 to simulations in case 1. The tested decreases on pixel-wise CCEs are negligible for all sectors, consistent with 
 the simulations in case1. The measured decreases on cluster-wise CCE for A0 and A7 are similar and are less than that of 
 large-pixel sectors , implying that large diode dimensions help on charge collection after irradiation with a constrained pixel 
 pitch (42\,$\mu$m in this work). The cluster-wise test results are also close to the simulations in case1.   

The test results demonstrate that \supixi sensor, suffering considerable loss of sensor performance especially on sensor gain and ENC,
 survives after neutron irradiation with fluence of $1 \times 10^{13}~1\,\text{MeV}~\text{n}_\text{eq}/\text{cm}^2$. 
 The pixels of all sectors keep almost the same CCE as that before irradiation and the effective SNR over 9. Besides,
 70\% of the CCE and 10 of the SNR, at least, are ensured for cluster. 
 Though not as excellent as pixels in A0, the large pixels are also potential for tracking on CEPC.
  In large pixels, small diodes, like those in A2, are preferred after irradiation for the best performance on SNR.

%%_____________________________________________________________________________end of article

\acknowledgments

This work has been supported by the National Natural Science 
Foundation of China(U1232202, U2032203 and 12075142), the 
Ministry of Science and Technology of China (2018YFA0404302) 
and Shandong Provincial Natural Science Foundation (ZR2020MA102).

%For bibtex
\bibliographystyle{JHEP_wm}
\bibliography{supix1_rad_MainTex}
 
\end{document}